\documentclass[preprint, prl]{revtex4}

\usepackage{graphicx}
\usepackage{dcolumn}
\usepackage{bm}

\begin{document}

\preprint{APS/123-QED}

\title{Charge ordering induces a smectic phase in oblate ionic liquid crystals}

\author{G. C. Ganzenm\"uller}
\email{georg.ganzenmueller@emi.fraunhofer.de}
\affiliation{Fraunhofer Ernst Mach Institute for High-Speed Dynamics,
Eckerstrasse 4, 79104 Freiburg, Germany}
\author{G. N. Patey,}
\email{patey@chem.ubc.ca}
\affiliation{Department of Chemistry, University of British Columbia, Vancouver, British Columbia, Canada V6T 1Z1}

\date{\today}

\begin{abstract}
We report a computer simulation study of an electroneutral mixture of
oppositely charged oblate ellipsoids of revolution with aspect ratio
$A= 1/3$. In contrast to hard or soft repulsive ellipsoids, which are
purely nematic, this system exhibits a smectic-A phase in which 
charges of equal sign are counterintuitively packed in layers perpendicular to the nematic
director.  
\end{abstract}

\pacs{Valid PACS appear here}
\maketitle

\section{\label{introduction}}  
While the phase behavior of uncharged mesogens, i.e. particles with anisotropic
shape, which give rise to liquid crystalline phase behavior, is now fairly
well understood \cite{Wilson:2005/a, Care:2005/a}, the generalization to
electroneutral mixtures of charged mesogens has received very little
attention. The existing computer simulation literature on ionic liquid
crystals \cite{Binnemans:2005/a} is mainly focused on so-called room-temperature ionic liquids
\cite{Castner:2010/a, Voth:2007/a}, and high-temperature melts of simple ionic crystals \cite{Spohr:2008/a}. 
The overwhelming majority of these studies considers real molecules, ranging from fully atomistic detail to
coarse-grained descriptions. There is a shortage of
investigations of fundamental systems, in which the shape anisotropy and the
Coulombic interactions can be described with as few parameters as possible.  An
important exception is the work by Avendano \textit{et al.}
\cite{Avendano:2008/a}, who studied binary mixtures of oppositely charged hard
spherocylinders.  While Avendano's study is of fundamental character because
spherocylinders are very well defined, it is desirable to employ a different
model which can continuously vary in shape between oblate and prolate forms. This way,
one can systematically investigate the structure as a function of particle
shape. We therefore consider charged ellipsoids of revolution, which are
uniquely defined by their aspect ratio, and which do not appear to have been previously studied.
The work presented in this Letter focuses on a single aspect
ratio at which very interesting phase behavior is obtained: charged oblate
ellipsoids with a height-to-width ratio $A=1/3$ feature a charge-ordered
smectic phase which has not been reported before. 

\begin{figure}
  \includegraphics*[width=\columnwidth]{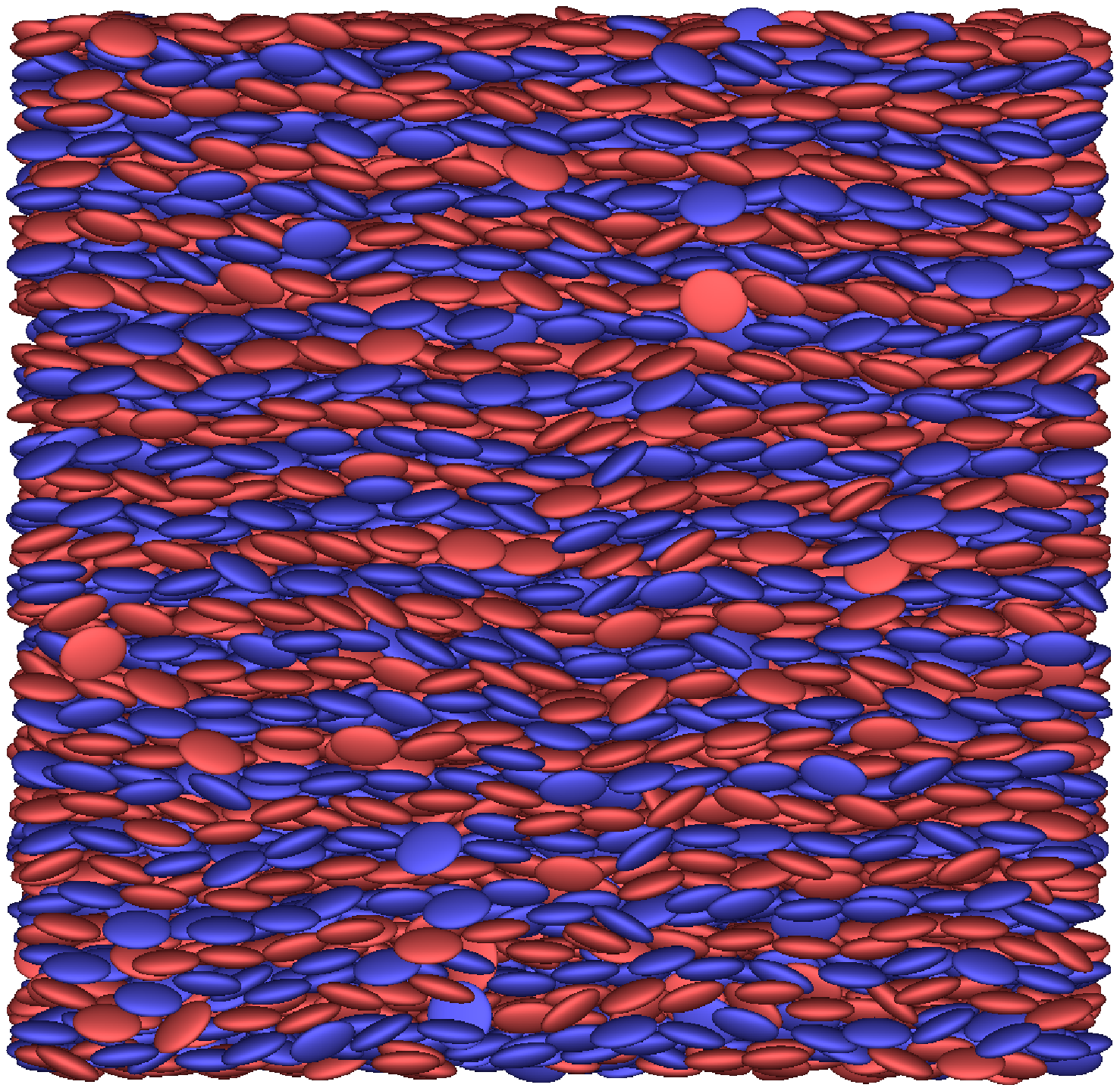}
  \caption{\label{fig:snapshot_layer}} Simulation snapshot of a system with
$N=8000$ particles at temperature $T^*=0.01$ and number density
$\rho^*=0.8$. Cations and anions are shown in red and blue.
Note that the direction of the charge layers is in general oriented randomly
with respect to the simulation box frame, but has been rotated here to
facilitate visualization of the charge ordering.
\end{figure}

\section{\label{details}
Model definition and computational details} 

We investigate electroneutral mixtures of $N$ oblate charged ellipsoids of revolution interacting via the
recently introduced anisotropic RE$^2$-pair potential of Everaers and Ejtehadi
\cite{Everaers:2003/a}, and Coulombic forces due to point charges located at
the particle's center.  The RE$^2$-pair potential originates from a
sophisticated overlap integral between two finite bodies each having a continuous density of 
sites, that interact via a
Lennard-Jones potential. In contrast to the widely used Gay-Berne potential
\cite{Gay:1981/a}, which is an approximation to a linear array of
Lennard-Jones particles, it reproduces both short- and long-range
interaction limits of ellipsoids of arbitrary shape with high accuracy, and has the additional advantage that it is well defined and has no adjustable parameters of unclear microscopic origin.
The RE$^2$-potential can be easily split into attractive and repulsive terms,
and it is only the latter which we employ here, as we wish to study the
interplay between anisotropic shape and electrostatic interactions. Instead of
quoting the lengthy expression for the orientation dependent potential here, we
refer the reader to Eq.~(36) of \cite{Everaers:2003/a}. The particle shape is
defined via a matrix $\mathcal{S} = \mathrm{diag}\{a,b,b\}$, where $a$ is the
short radius along to the axis of revolution, and $b$ is the conjugate radius
defining the lateral elongation. Because we retain only the repulsive part of the
Everaer-Ejtehadi potential, it becomes necessary to parameterize the spatial extension
of the ellipsoidal particles in order to define the length scale of the the
system. We chose to keep the original parameters and scale the shape-matrix
$\mathcal{S}$ with an additional factor $s$, which is chosen such that, in the
special case of a spherical particle with $\mathcal{S} =
s\times\mathrm{diag}\{\frac{1}{2},\frac{1}{2},\frac{1}{2}\}$, the value of the
repulsive pair potential attains $1\epsilon$ at a particle separation of
$1\sigma$. The required value $s\simeq0.35492675\sigma$ was determined
numerically. Defining the aspect ratio $A=a/b$, we chose $a =
\frac{1}{2}A^{2/3}$ and $b = \frac{1}{2}A^{-1/3}$, which fixes the occupied
volume fraction, irrespective of the aspect ratio, to be the same as that of an
equivalent system of ``unit spheres'' with $A=1$ \cite{method:EllipsoidVolume}.
Here, we study mildly oblate particles with $A=\frac{1}{3}$. The length and
energy scales, $\sigma$ and $\epsilon$, are henceforth used to quote all results
in the usual set of reduced units \cite{Allen:1987/a}: number density
$\rho^*=N/V\sigma^3$, temperature $T^*=k_B T/\epsilon$, energy
$E^*=E/\epsilon$, pressure $p^*=p\sigma^3 / \epsilon$, and time
$\tau=t\sqrt{\epsilon/m\sigma^2}$ with $m=1$.

Simulations were carried out with the MD package LAMMPS \cite{Plimpton:1995/a},
suitably modified to incorporate our specific potential. The RE$^2$-potential
was cut off at $2\sigma$. Coulomb interactions were handled using both Ewald
\cite{Perram:1988/a} and Particle-Particle Particle-Mesh \cite{Hockney:1981/a}
summation techniques with conducting boundary conditions, and real-space cutoff
of $5\sigma$. The system size for the numerical results reported here is $N=1000$,
however, we have carefully checked system size dependencies up to $N=32768$ and
found no significant difference. In addition, we cross checked the validity of
the MD results with Monte Carlo simulations using the Wolf method \cite{Wolf:1999/a} to handle
Coulomb interactions, obtaining quantitative agreement. Because the Wolf method
is a direct summation technique with no long-range correction, we can thus rule
out artificial stabilization of the observed phases via the reciprocal space
long-range correction. The equations of motion were integrated with the
Velocity-Verlet scheme using a reduced time step $\delta\tau=0.0075$ in the
$NVT$-ensemble with Nose-Hoover thermostatting.  In order to accurately
determine transition temperatures between different phases, a careful annealing
procedure was followed: initial configurations were chosen randomly at a
dilute concentration of $\rho^{*}=0.01$, and the system was slowly adjusted
during $10^4\tau$ at $T^*=0.3$ to the desired target number density
$\rho^*=0.8$ (corresponding to an occupied volume fraction $\phi\simeq42\%$) by
isotropically decreasing the box boundaries every time step by an amount
proportional to the box volume. Subsequently, the temperature was gradually
decreased in intervals of $\delta T^*=0.005$ over $2000\tau$, followed by a
further equilibration period of $2000\tau$. Data for analysis was then
collected during a production run of length $5000\tau$. The total number of
simulation time steps required for all temperatures was thus $\simeq
53\times10^6$.

\section{\label{Results} Results}
\begin{figure}
  \includegraphics*[width=\columnwidth]{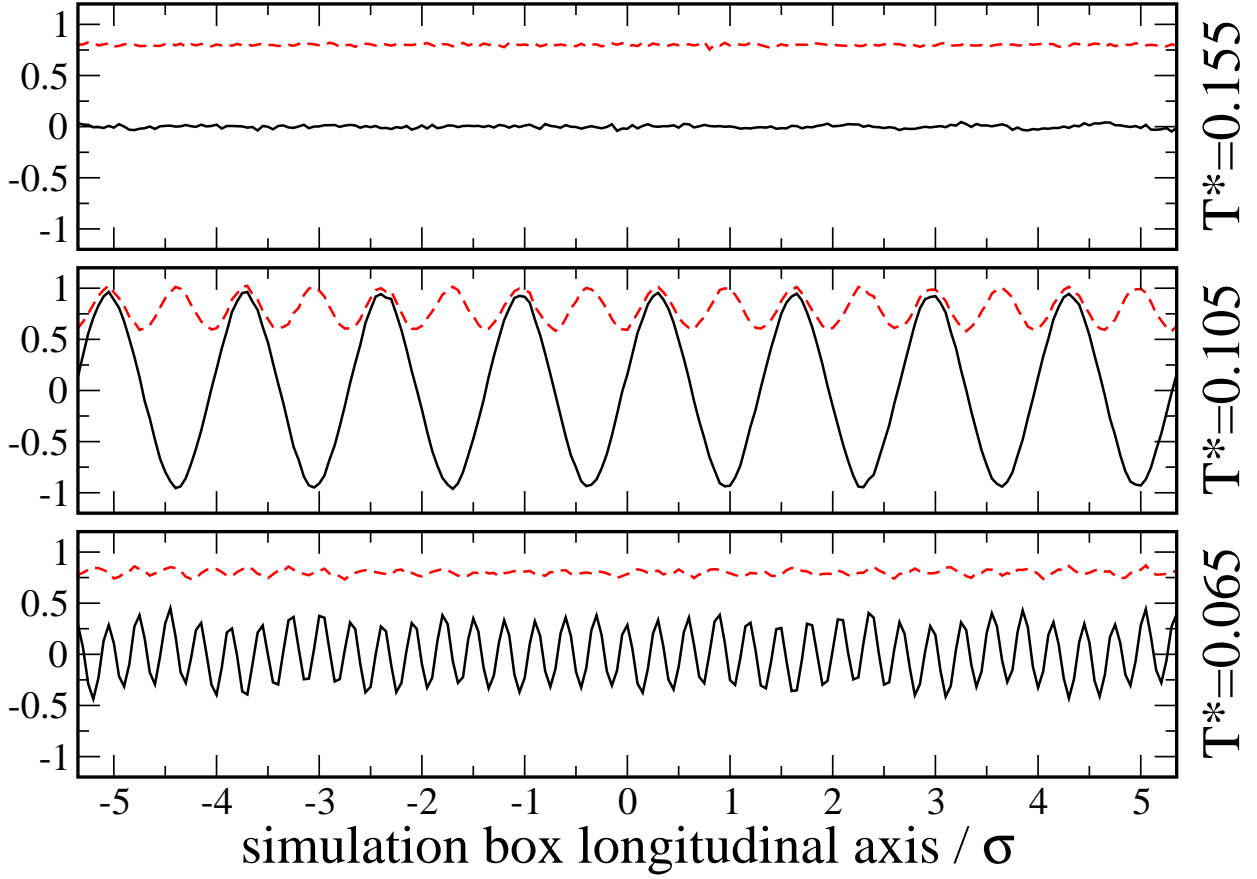}
  \caption{\label{fig:profiles}
Mass (dashed lines) and charge (solid lines) density profiles corresponding
to the three different phases of oblate charged ellipsoids discussed here. Top:
High temperature nematic phase with no charge ordering. Center: Smectic,
charge-ordered phase. Bottom: Hexagonal columnar low temperature phase. The
longitudinal axis of the systems is the Eigenvector associated with $P_2$
for both the nematic and the smectic phase, while for the low-temperature
columnar phase the dipole-director associated with $P_2^d$ is used (see text). The
mass (charge) density profile is calculated by summing over all masses
(charges) located within a slab of width $0.1\sigma$ normal to the longitudinal
axis and dividing by the slab volume.}
\end{figure}

\begin{figure}
\includegraphics*[width=\columnwidth]{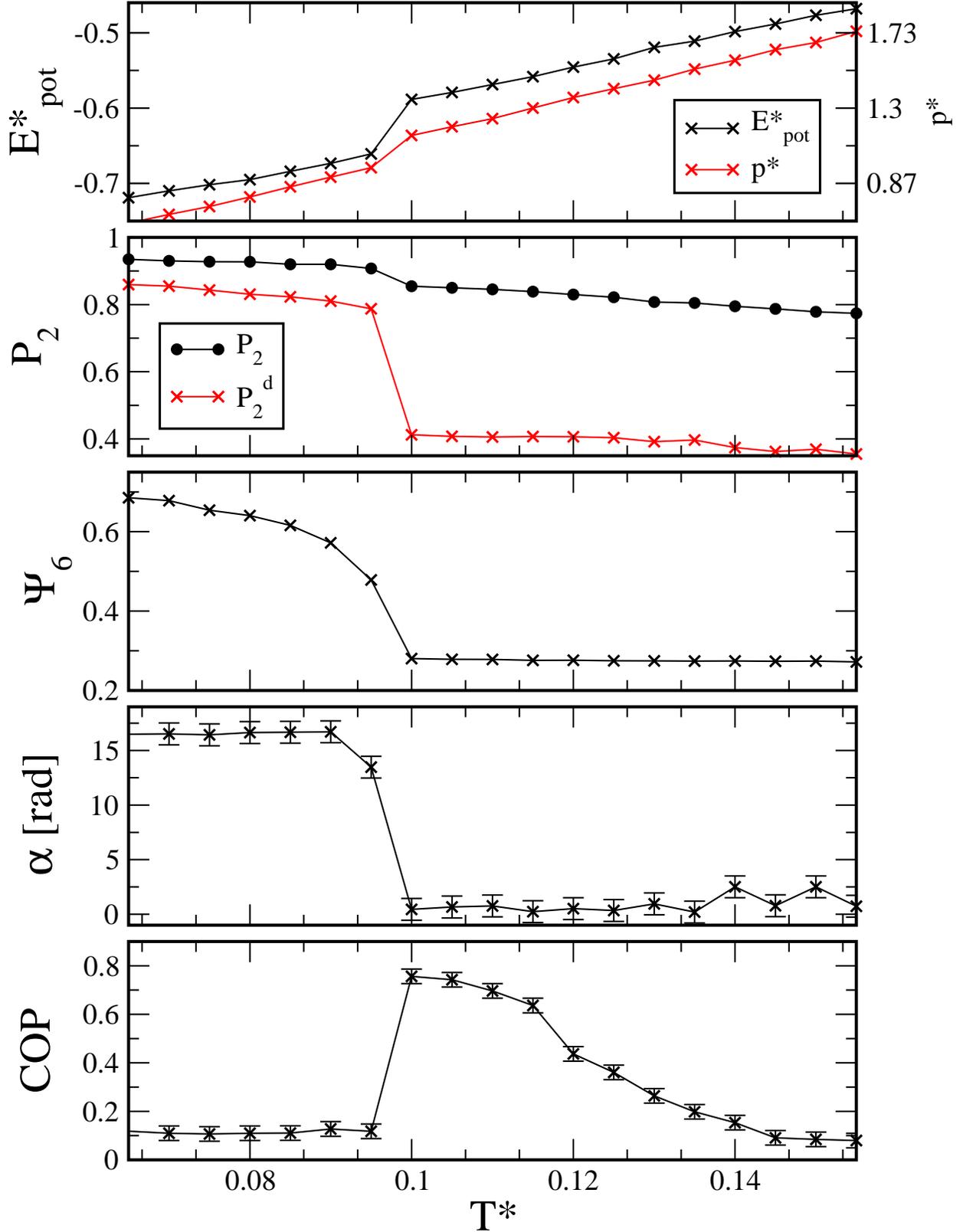}
\caption{\label{fig:ops} Results for the order parameters (defined in the text)
as function of temperature. Error bars are not shown if the symbol size is
not exceeded. Note the two different ordinates for pressure and potential energy in
the top graph.}
\end{figure}

At the number density considered here, the system exhibits a finite value of
the nematic order parameter $P_2$ across the temperature range considered, 
$T^*=0.03-0.3$.  $P_2$ is defined as the largest eigenvalue of the ordering
tensor $S_{\alpha\beta} = \frac{1}{N} \sum_{i=1}^N \frac{1}{2} (3 e_{i\alpha}
e_{i\beta }- \delta_{\alpha \beta}$), where $\alpha, \beta = x, y, z$ are the
indices referring to the space fixed frame, $\delta_{\alpha \beta}$ is the
Kronecker delta, and $e_{i\alpha}$ is a component of the ellipsoid orientation
vector, taken to be the major axis along which the short radius $a$ is defined.
Above $T^*\simeq0.13$, the system is purely nematic, with no long-range mass or
charge ordering apparent. Below this temperature, a charge-ordered, smectic
phase emerges, see Fig.~\ref{fig:snapshot_layer}.  Interestingly, and quite
counterintuitively, particles of the same charge are located in layers. The
nematic director (the eigenvector associated with $P_2$) is normal to these
planes, thus defining the longitudinal axis of this phase. Na\"ively, one could
have expected a temperature-driven transition between a positionally disordered
phase and a solid columnar phase, where stacks of alternatingly charged
discoids are shifted parallel to each other, such that opposite charges are
found in the transverse direction. However, we find that electroneutral
mixtures of charged oblate particles feature an intermediate phase, where
smectic ordering is realized in conjunction with charge ordering. Both the
smectic character of this phase and the charge ordering have been verified by
calculating the mass $\rho_m(z)$ and charge $\rho_q(z)$ density profiles along
the longitudinal direction normal to the charge layers, see
Fig.~\ref{fig:profiles}. While $\rho_m(z)$ and $\rho_q(z)$ show no features at
temperatures $T^* \gtrsim 0.13$, both charge and mass ordering are evident in
the temperature range $0.1 \lesssim T^* \lesssim 0.13$, clearly identifying a
smectic phase in which mass and charge ordering are strongly coupled. At
temperatures $T^* \lesssim 0.1$, a highly ordered phase emerges, which appears
to be a crystalline solid, as will be discussed below.  We note at this point
that the same system \textit{without charges}, i.e. interacting only via a
repulsive potential, is nematic between $T^*=0.3$ to $T^*\simeq0.04$, and
freezes into a crystalline solid upon further cooling. This is in agreement
with theoretical considerations that hard ellipsoids do not
feature a liquid smectic-A phase \cite{Evans:1992/a}. We thus conclude that
the smectic phase we observe is induced by the Coulombic interactions. In
order to quantify the degree of charge ordering, we define a charge-order
parameter $COP = \int_{-L/2}^{L/2} |\rho_q(z)| \mathrm{d}z / \int_{-L/2}^{L/2}
|\rho_m(z)| \mathrm{d}z$, where $L$ is the length of the simulation box.  $COP$
attains a value close to zero for the disordered high-temperature phase and
reaches values close to unity at $T^*=0.1$, see Fig.~\ref{fig:ops}. Below this
temperature, in the crystalline solid, the order parameter reduces to $COP
\simeq 0.2$. This can be explained using the picture introduced above of the
stacks of alternatingly charged discs which are shifted parallel to each other,
such that particles of opposite charge are also found laterally: the shifting
destroys the charge layers. 

In order to quantify the correlation between ion pairs, it is instructive to
consider an alternate set of local directions present in the system. For each
particle, we identify the spatially closest neighbor of opposite charge. The
normalized distance vector between these two particles can be regarded as a
local dipole vector, and we calculate the largest eigenvalue $P_2^d$ and
associated dipole director of the ordering tensor $S_{\alpha \beta}$ using
these orientations.  Analyzing these quantities, we find that there is no
enhanced dipole correlation in the charge-ordered phase over the
high-temperature nematic phase, see Fig.~\ref{fig:ops}. However, upon further
cooling below $T^*=0.095$, $P_2^d$ and $P_2$ increase suddenly, indicating the
formation of long-range discoid stacks of alternatingly charged particles. This
marks the transition to the low-temperature hexagonal columnar phase, which one
could have expected na\"ively to be the minimum energy configuration. This
transition appears to be of first order, as evidenced by the discontinuities
(which of course suffer from finite-size rounding effects) in both pressure $p^{*}$
and potential energy $E^{*}_{pot.}$

\begin{figure}
  \includegraphics*[width=\columnwidth]{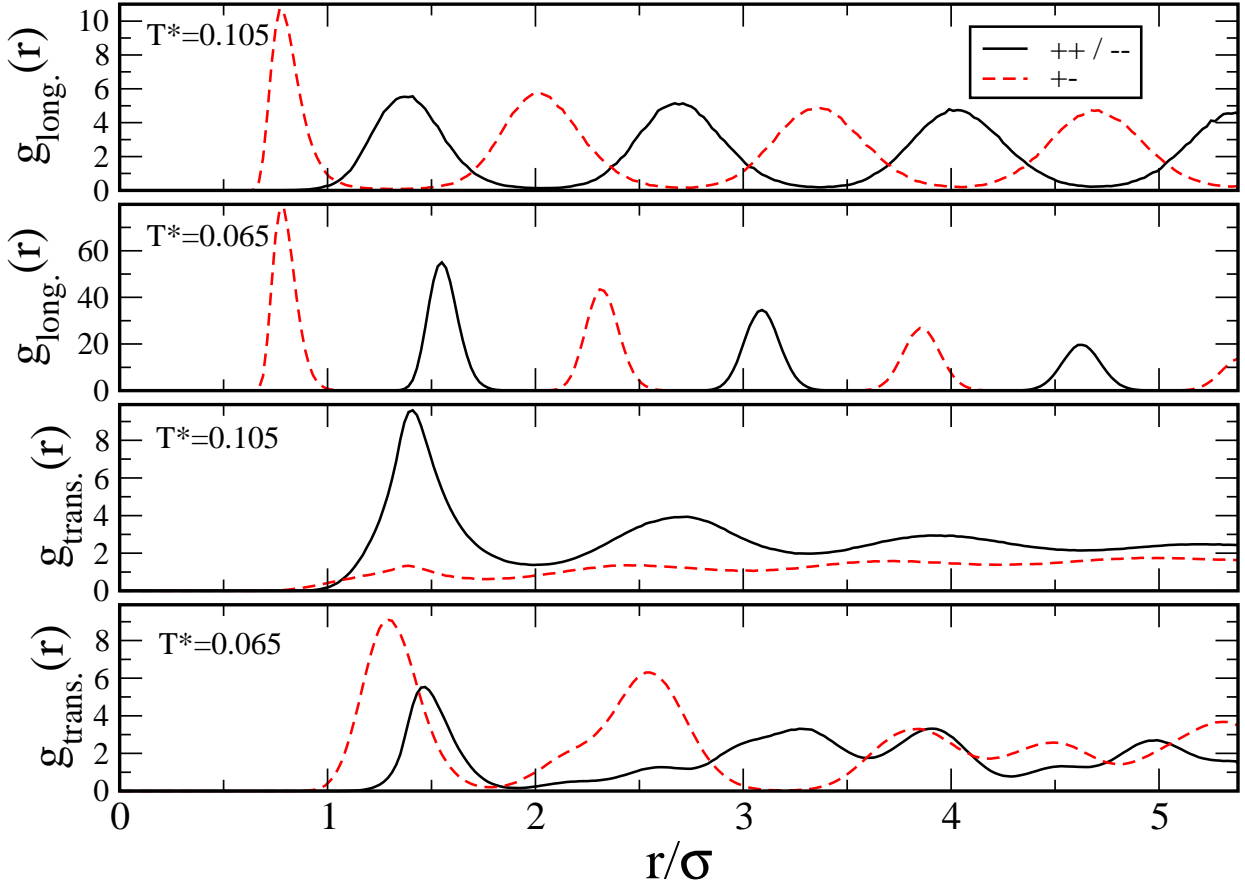}
  \caption{\label{fig:RDF}
  Charge correlation functions $g_{long.}(r)$ and $g_{trans.}(r)$ along the
longitudinal and transverse directions, respectively. Results are shown for the
liquid-crystalline charge-ordered phase at $T^*=0.105$, and the solid
crystalline phase at $T^*=0.0605$. $++ / --$ denotes correlation functions
between particles of the same charge, and $+-$ denotes the correlation
functions for pairs of opposite charge.}
\end{figure}

The hexagonal columnar character of the low-temperature phase is confirmed by
the hexagonal bond-order parameter $\Psi_6$ \cite{method:Phi6},
which also increases suddenly near the transition temperature $T^*\simeq0.095$.
The aforementioned shifting of the columns relative to each other, such that
charges of opposite sign are also found along the transverse direction, is
evidenced by the in-plane radial distribution functions $g_{trans.}(r)$ shown
in Fig.~\ref{fig:RDF}. At $T^*=0.105$, the system is charge-ordered smectic,
and the in-plane probability of finding a neighbor of the same charge is always
higher than the probability of finding a neighbor with opposite charge. This
situation is reversed at $T^*=0.065$, where the system shows columnar ordering.

Interestingly, the columnar phase appears to be tilted, i.e., the orientation
of the oblate particle's major axis does not coincide with the column axis. It
has been recently observed \cite{Chakrabarti:2008/a} that attractive
interactions induce the formation of tilted columnar phases  in discotic
mesogens. Here, we identify the columnar axis with the dipole director defined
above, because, the local interparticle distance vector between spatially
closest ion pairs points necessarily along the stacking direction for flat
objects like the oblate particles considered here. The angle between the nematic
and dipole directors increases from zero to a tilt value of $\alpha \simeq
16^{\circ}$ as the system is cooled from the liquid-crystalline smectic phase
to the crystalline hexagonal columnar phase.

\section{Discussion}
We have investigated an electroneutral symmetric mixture of soft repulsive
oblate ellipsoids with point charges located at the particle's center. At an
occupied volume fraction $\phi\simeq42\%$ , we find a high-temperature nematic
phase which is also known to exist in liquids of uncharged oblate mesogens
\cite{Frenkel:1982/a}. In case of the symmetric Coulombic system considered
here, we report, for the first time, a charge-ordered smectic
liquid-crystalline phase which emerges at lower temperatures close to the
freezing temperature. In this phase, particles of equal charge are organized in
layers which are normal to the nematic director, such that smectic- and
charge-ordering are invariably coupled. The layered arrangement yields a
reasonably low energy and, at the same time, affords a relatively high
configurational entropy because there is no positional order within a layer. It
is this balance between enthalpy and entropy which maintains the favorable free energy,
and thus stabilizes this peculiar phase.  Upon further cooling, we
observe a transition to a solid columnar phase which is accompanied by a shift
along the column axis. The shifted phase realizes an even lower energy, but at
the same time reduces entropy by enforcing positional order perpendicular to
the columnar axis.  The stability of the layered phase has been checked using
large systems with up to 32768 particles, in order to rule out artificial
stabilization due to periodic boundary effects. We note however, that for
larger systems even slight undulations of the charge layers and dislocation
defects (which are unavoidable due to entropic stabilization) will eventually
destroy the perfect charge-ordering.  Nevertheless, the properties, particularly the conductivity,
of this phase promise to be interesting candidates for future studies, as the
screening lengths parallel and orthogonal to the layer directions must be
very different. Recent developments of colloidal synthesis techniques could make an
experimental confirmation of this charge-ordered phase possible, as it is now
possible to produce mesogens of continuously varying aspect ratio
\cite{Snoeks:2000/a} as well as stable mixtures of oppositely charged colloids
\cite{Yethiraj:2003/a, Royall:2003/a}.

\begin{thebibliography}{22}
\bibitem[{met({\natexlab{a}})}]{method:EllipsoidVolume}
\bibinfo{note}{The volume of an ellipsoid of revolution is $V=\frac{4}{3}\pi a
  b b$, which becomes $V=\frac{4}{3}\pi (\frac{1}{2})^3$ with $a =
  \frac{1}{2}A^{2/3}$ and $b = \frac{1}{2}A^{-1/3}$ for all $A$.}
\bibitem[{met({\natexlab{b}})}]{method:Phi6}
\bibinfo{note}{The hexagonal bond order parameter is defined as $\Psi_6=\left|
  \langle \frac{1}{N} \sum_j \frac{1}{n_b^j} \sum_{\langle kl \rangle} w_{kl}
  \exp(6i\theta_{kl}) \rangle \right|$, where $n_b^j$ is the number of pairs of
  nearest neighbors of the $j$th particle (obtained via Voronoi tesselation),
  the inner sum goes over all possible pairs of neighbors, and $\theta_{kl}$ is
  the angle between the unit vectors along the projections of the
  intermolecular vectors between particle $j$ and its neighbors $k$ and $l$
  onto a plane perpendicular to the dipole director, which represents the
  columnar axis. The preexponential factor $w_{kl}$ was taken to be unity if
  the separation vectors $\mathbf{r}_{jk}$ and $\mathbf{r}_{jl}$ lie within a
  hollow cylinder of height $1.2\sigma$, outer radius $L$, and inner radius
  2.2, centered at particle $j$, thus effectively excluding neighbors belonging
  to the same column, and zero otherwise.}
\end{thebibliography}
\end{document}